\newcounter{myctr}
\def\myitem{\refstepcounter{myctr}\bibfont\noindent\ifnum\themyctr>9\else\phantom{0}\fi\hangindent17pt\themyctr.\enskip}

\documentclass{ws-ijqi}

\newcommand{\mH}{\mathcal{H}}
\newcommand{\mM}{\mathcal{M}}
\newcommand{\mT}{\mathcal{T}}

\newcommand{\mZ}{\mathcal{Z}}

\newcommand{\ep}{\epsilon}

\begin{document}

\markboth{A. Abidin, J-A Larsson}
{Vulnerability of ``A novel protocol-authentication algorithm\ldots}

\catchline{}{}{}{}{}

\title{VULNERABILITY OF ``A NOVEL PROTOCOL-AUTHENTICATION ALGORITHM\\
RULING OUT A MAN-IN-THE-MIDDLE ATTACK\\
IN QUANTUM CRYPTOGRAPHY''   }

\author{AYSAJAN ABIDIN and JAN-\AA KE LARSSON}

\address{Department of Mathematics, Link\"oping University\\
581 83 Link\"oping, Sweden\\
E-mail: aiabu@mai.liu.se and jalar@mai.liu.se}

\maketitle


\begin{abstract}
  In this paper we review and comment on ``A novel
  protocol-authentication algorithm ruling out a man-in-the-middle
  attack in quantum cryptography'', [M. Peev {\it et al}., {\it Int.
    J. Quant. Inform.}, {\bf 3}, 225, (2005)]. In particular, we point
  out that the proposed primitive is not secure when used in a generic
  protocol, and needs additional authenticating properties of the
  surrounding quantum-cryptographic protocol .
\end{abstract}


\section{Introduction}

Quantum Cryptography, or more accurately Quantum Key Distribution
(QKD), is an unconditionally secure key growing technique based on the
principles of quantum mechanics. It is unconditionally secure because
no quantum state can be copied or measured without disturbing it.
However, the practical implementation of QKD protocols requires an
immutable public channel. In case the public channel is not immutable,
the eavesdropper (Eve) can easily mount a man-in-the-middle (MITM)
attack, since Eve is in control of both the quantum and the public
channels. For the attack to be successful Eve needs, among other
things, to substitute the classical message from one legitimate user
(Alice) to the other (Bob) without being noticed.  To prohibit such an
attack on QKD, proper message authentication is needed. Therefore, QKD
is secure only if it is combined with an unconditionally secure
message authentication scheme.  In this paper we will review a
recently proposed authentication primitive\cite{1} and point out that
it is not secure when used in a generic QKD system. It has earlier
been shown\cite{BMQS} that an attack is possible against the ``privacy
amplification'' step in a QKD protocol using the proposed
authentication, but the attack presented here is more serious and
enables a full MITM attack on the whole system, unless some additional
part of the protocol has authenticating properties.

\section{The proposed authentication primitive}

In Ref.\ \refcite{1}, the authors propose an authentication primitive
which aims at decreasing the key consumption for the authentication
purposes in QKD, and in turn to improve the efficiency of the key
growth in QKD. The algorithm works as follows. Let $\mM$ be the set of
all binary strings of length $m$ (or the set of all messages of length
$m$), and let $\mT$ be the set of all binary strings of length $n$
with $n<m$ (or the set of all tags of length $n$). A message $m_\text
A$ is first mapped from $\mM$ to $\mZ$, where $\mZ$ is the set of all
binary strings of length $r$ with $n<r<m$, by a single publicly known
hash function $f$ so that $z_\text A = f(m_\text A)$. And then,
$z_\text A$ is mapped by a secret $h_k\in\mH_{\mZ}$ to a tag $t_\text
A=h_k(z_\text A)$, where $\mH_{\mZ} : \mZ \mapsto \mT$ is a Strongly
Universal$_2$ (SU$_2$) family of hash functions\cite{WC} and the subscript 
$k$ is the secret key needed to identify a hash function. The message-tag
pair $m_\text A+t_\text A$ will be sent over the public channel. To
authenticate the message $m_\text A\in\mM$, the legitimate receiver
computes $h_k(f(m_\text A))$ and compares it to $t_\text A$. If they
are identical then the message will be accepted as authentic,
otherwise it will be rejected. Since $r$ is fixed independently of
$m$, the key length required for authentication is constant regardless
of the message length to be authenticated.

This authentication algorithm is claimed\cite{1} to be secure with a
probability $\ep$ of Eve being able to create the correct tag for her
fake message. In Ref.\ \refcite{1}, this is calculated
as\footnote{Actually, $\ep \le \ep_1 + \ep_2$; eqn.\ (\ref{eq:1}) is
  an upper bound rather than an equality.}
\begin{equation}\label{eq:1}
  \ep = \ep_1+\ep_2
\end{equation}
where $\ep_2 = 1/|\mT|$ which is the probability of guessing the
correct tag when a SU$_2$ hash function family is used and $\ep_1$ is
the probability that the message $m_\text A$ and Eve's modified
message $m_\text E(\neq m_\text A)$ yield the same value under the
publicly known hash function $f$.

\section{The problem}

This authentication primitive is such that whenever Eve's message
$m_\text E$ happens to coincide with Alice's message $m_\text A$ under
the publicly known hash function $f$, i.e. $f(m_\text E)=f(m_\text
A)$, Eve can just send $m_\text E + t_\text A$ since $t_\text E =
t_\text A$. The problem here is that in Ref.\ \refcite{1} security is
derived under the explicit assumption that Eve has a fixed message.
The result holds, but in generic QKD Eve is \emph{not} restricted to
one message $m_\text E$.

In a full MITM attack on a QKD protocol, Eve impersonates Bob to Alice
and Alice to Bob during the quantum transmission process and the
subsequent public discussions.  We use BB84\cite{BB84} with simple
reconciliation and privacy amplification; and \emph{immediate
  authentication} of each phase as our first example. This would
consist of, in order, raw key generation; sifting and immediate
authentication; one-way error correction and immediate authentication;
one-way privacy amplification and authentication (see, e.g., Ref.\
\refcite{NL} Chapter 12). Eve receives and measures the qubits that
Alice has sent to Bob, in her choice of basis. We note here that
although QKD requires that Bob randomly selects the basis to measure
the qubits in, Eve can ignore this requirement. At the same time she
chooses a set of qubits in, again, not necessarily random states and
sends these to Bob. After Bob receives and measures the qubits sent by
Eve in a randomly selected basis, he sends an authenticated time stamp
to Alice to end the quantum transmission phase. Now Alice sends her
message $m_\text A$, which contains the settings used for
encoding/decoding on the quantum channel, along with the
authentication tag $t_\text A$ to Bob. Eve intercepts the message-tag
pair and calculates $f(m_\text A)$ and compares it with $f(m_\text
E)$. In the rare event that they are equal, Eve can just send $m_\text
E+ t_\text A$ to Bob. Otherwise, she can change her message $m_\text
E$ which contains the settings. Changing one of the settings, i.e.,
changing one bit of the message, will at most introduce one noisy bit
in the sifted key.  Even a few noisy bits will not make a noticeable
effect in practical QKD systems because of the error correction used
in the reconciliation step.

In this situation, if $f(m_\text E)\neq f(m_\text A)$, Eve can search
for a message $m_\text E'$ with $d_\text{Hamming}(m_\text E, m_\text
E') = 1$ (or ``small'') such that $f(m_\text E') = f(m_\text A)$. In
other words, she tries to find a collision between $m_\text A$ and
$m_\text E'$ under $f$ such that $m_\text E'$ is close to $m_\text E$,
and it is well known that such collisions may exist for many hash
functions and in fact do exist for well-known examples\cite{2,3}.  Eve
can now send the message-tag pair $m_\text E'+t_\text A$ knowing that
Bob will accept the message $m_\text E'$ as authentic.

Searching for a collision requires Eve to have sufficient computing
power, but usually in QKD no bounds are assumed on Eve's computing
power. One should also note that the computing power needed may be
lower than one would first expect\cite{2,3}. However, even without
sufficient computing power, Eve can make a list of different values of
$m_\text E'$ and the corresponding value of $z_\text E' = f(m_\text
E')\in\mZ$ in advance, and save it in her device. Remember that the
usual requirement of having random settings (making the message
$m_\text E$ random) does not apply to Eve; the requirement is needed
to ensure that the final key is secret, something that Eve can ignore.
With a pre-chosen $m_\text E$, a list of pairs $(m_\text E',z_\text
E')$ and her received $m_\text A+t_\text A$, Eve can just compute
$z_\text A = f(m_\text A)$ and pick $m_\text E'$ from her list
corresponding to $z_\text A$, and then send $m_\text E' + t_\text A$.
She can even make a partial list, and simply wait for the first match
to occur. In fact, the parameter $\epsilon_1$, now interpreted as the
probability that some item in Eve's list collides with $m_\text A$,
depends linearly on the size of this list.  If she is able to make a
full list (one message $m_\text E'$ for each possible $z_\text A$), or
has sufficient computing power, she is certain of success in the
sifting phase every time she performs the MITM attack.

Eve now has two sets of sifted keys, one shared with Alice and the
other with Bob. The remaining steps are one-way error correction and
authentication; and one-way privacy amplification and authentication.
These are completed by sending random parity maps over the classical
channel, and in case of error correction also the parity values
\cite{BBBSS,BS,BBR,BBCM}.  In the case of error correction, Eve
intercepts the authenticated error-correction information (random maps
and the output values) sent by Alice to Bob, and error-corrects the
sifted key that she shares with Alice.  She then searches for
\emph{non-random} maps (and corresponding output) of the sifted key
shared with Bob, that makes her message collide with Alice's under
$f$. Note that Eve at this point may change any bit of the sifted key
at the price of introducing an extra bit error in the sifted key. This
will enable a collision even if all the possible maps do not. She
sends the resulting message to Bob along with Alice's tag, which will
then be accepted by Bob.  Bob responds by an authenticated message
that signals which subsets matched and which subsets were successfully
error-corrected, and also indicates the error rate of the sifted key;
in this simple scheme this is used as error estimate.  Eve modifies
her corresponding but still waiting response to Alice so that it will
collide with Bob's message under $f$. This may introduce some noise
into the error-corrected key shared between Alice and Eve, but this
goes unnoticed by Alice unless an extra detection phase is present
(see below).

The privacy amplification is performed by Alice choosing a random map,
and sending that over the classical channel, whereafter Alice and Bob
apply this map to their respective reconciled keys.  Here, Eve
intercepts the description of the map and the tag, and privacy
amplifies the reconciled key (shared with Alice) using the received
map.  She then searches for a new \emph{non-random} map to use for
privacy amplification with Bob that makes the message coincide with
Alice's under $f$. If Eve arranges for the reconciled key shared with
Alice to be of equal length to that shared with Bob, she can even
reuse the map that Alice sent.  Then, Eve sends the chosen map along
with Alice's tag to Bob, who will accept them and privacy amplify his
error-corrected key accordingly.

\section{Countermeasures}

The situation is improved if postponed authentication is used, or for
example, when using iterative reconciliation methods.  More precisely,
if the messages are sent in each phase as usual (sifting, error
correction and privacy amplification, etc.) but not authenticated
until the end of the round, then Eve's freedom to change her message
is restricted to the message part in the last phase.  And this
severely restricts Eve's possibilities, even though an attack is still
possible as is shown in Ref.\ \refcite{BMQS}.

Another more effective improvement is to use secret key in an
additional phase of the protocol. There is no explicit mention of
using more secret key for this purpose in Ref.\ \refcite{1} but it is
implicit; it is present in their reference 5 (here Ref.\
\refcite{GH}).  The procedure basically uses already shared secret key
to choose a hash function to detect errors in the reconciled key.
Another suggestion is to one-time pad the reconciliation
procedure\cite{Lutken}.  Both of these suggestions are intended to
keep the information leaked in error correction at a minimum, but they
also implicitly add an authentication property of that phase.  Using a
modification like this will probably improve the situation but the
needed formal proof is beyond the scope of this paper. It is perhaps
important to note that this puts stronger requirements on the extra
cryptographic primitives used since they are used as authentication in
addition to limiting the information leakage. But, since the mentioned
modifications both use cryptographically secure primitives, it is to
be expected that they are resilient to extra demands of this type.

\section{Conclusion}

This brief review of a proposed authentication algorithm intended to
rule out a man-in-the-middle attack in QKD shows that the proposed
method is insecure when used in a generic QKD protocol. The main
problem is that Eve is not limited to a fixed (random) message, but
can in fact choose what message to send, and can check if her chosen
message gives the same tag as Alice's message, since the first-step
hash function $f$ is publicly known.

Using extra shared secret key for an extra authentication in one of
the phases probably improves the situation, but it should be stressed
that, unlike Wegman-Carter authentication, the security of the
proposed authentication procedure is highly dependent of the context
in which the authentication is applied.

Therefore, in general, great care should be taken when authentication
primitives used in the context of QKD are not
information-theoretically secure.

\end{document}